\documentclass[11pt,twoside]{article}

\usepackage{asp2006}
\usepackage{epsf}
\usepackage{psfig}
\usepackage{lscape}

\begin{document}
\title{AGN astrophysics via multi-frequency monitoring of $\gamma$-ray blazars in the {\sl Fermi}-GST era}   %%% Fill in title
%\author{E. Angelakis}   %%% Fill in author names
%\affil{Max-Planck-Institut fuer Radioastronomie, Auf dem Huegel 69, Bonn 53121, Germany}    %%% Fill in author affiliations
\author{E. Angelakis\altaffilmark{1}, L. Fuhrmann\altaffilmark{1}, J. A. Zensus\altaffilmark{1}, I. Nestoras\altaffilmark{1}, N. Marchili\altaffilmark{1}, T. P. Krichbaum\altaffilmark{1}, H. Ungerechts\altaffilmark{2}, W. Max-Moerbeck\altaffilmark{3}, V. Pavlidou\altaffilmark{3}, T. J. Pearson\altaffilmark{3}, A. C. S. Readhead\altaffilmark{3}, J. L. Richards\altaffilmark{3} and M.~A.~Stevenson\altaffilmark{3}}   %%% Fill in author names
\altaffiltext{1}{Max-Planck-Institut f\"ur Radioastronomie, Bonn 53121, Germany}
\altaffiltext{2}{Instituto de RadioAstronom\'ia Milim\'etrica, Granada E 18012, Spain}
\altaffiltext{3}{California Institute of Technology, Pasadena, CA 91125, USA}

\begin{abstract} %%% Abstract to run on from here.
The {\sl F-GAMMA}-project is the coordinated effort of several observatories to understand the AGN phenomenon and specifically blazars via multi-frequency monitoring in collaboration with the {\sl Fermi}-GST satellite since January 2007. The core observatories are: the Effelsberg 100-m, the IRAM 30-m and the OVRO 40-m telescope covering the range between 2.6 and 270 GHz. Effelsberg and IRAM stations do a monthly monitoring of the cm to sub-mm radio spectra of 60 selected blazars whereas the OVRO telescope is observing roughly 1200 objects at 15 GHz with a dense sampling of 2 points per week. The calibration uncertainty even at high frequencies, is of a few percent. 47\% of the Effelsberg/Pico Veleta sample is included in the LBAS list. An update of the monitored sample is currently underway.  \end{abstract}

%%% MAIN BODY OF TEXT GOES HERE. CONSULT "INSTRUCTIONS FOR AUTHORS USING
%%% LATEX2E MARKUP", SECTIONS 2.3-2.6 FOR HELP WITH EQUATIONS, FIGURES,
%%% AND TABLES.

\section{Introduction}   
The {\sl unified model} paradigm for the Active Galactic Nuclei (AGN), attributes the wealth of AGN classes mostly to different parameters of the same principal system viewed at different angles of the line-of-sight to the relativistic jet. Radiogalaxies and blazars are two extremes: in the former the jet is seen from large whereas in the latter from small angles ($\theta\le 20^\circ$).

Blazars are dominated by relativistic beaming and they exhibit intense variability at all wavelengths and time scales, highly superluminal apparent speeds, a significant degree of fractional polarization and polarization variability. Shock-in-jet  \citep[][]{marscher1985, hughes1985, marscher1996}, colliding relativistic plasma shells \citep{guetta2004}, lighthouse effect \citep{camenzind1992} or MHD-instabilities in the accretion disks \citep{begelman1980,villata1999}  are some of the models that attempt to explain variability at different time scales. Although the exact physical processes at play are unclear, the study of the temporal behavior of the SED can shed light on emission mechanism since different mechanisms predict different variability patterns \citep[e.g.][]{bottcher2002}. Hence, multi-frequency monitoring of blazars is essential in understanding the blazar physics.
   
\section{Current status, results and discussion}
{\bf The F-GAMMA alliance: }The Large Area Telescope (LAT) on-board {\sl Fermi}-GST offers a unique opportunity covering the $4\pi$ sky every three hours providing $\gamma$-ray light curves at $\sim$30 MeV - 300 GeV. To exploit that, we have initiated a tightly coordinated ``alliance'' of teams \citep[F-GAMMA project,][]{angelakis2008, fuhrmann2007} in order to understand the AGN physics via multi-frequency monitoring. The core program involves the 100-m MPIfR telescope (Effelsberg, Bonn, 2.64 - 43.00 GHz), the 30-m IRAM telescope (Pico Veleta, Granada, 86, 142, 220 and 270 GHz) and the 40-m OVRO telescope (Owens Valley, California, 15 GHz). The Effelsberg and the IRAM telescopes do a monthly monitoring of $\sim$60 blazars compiled on the basis of their $\gamma$-ray detectability \citep[EGRET][]{hartman1999} and their presence in the ``high priority list'' of the LAT AGN group. The OVRO telescope is monitoring a statistically complete sample of 1200 sources based on the CGRaBS catalog \citep{healy2008}. The sampling rate is 2-3 times a week \citep[see][]{richards2009,maxmoerbeck2009}.   
\\{\bf The first 2.5 years:} The F-GAMMA project has been running since January 2007 at Effelsberg and June 2007 at OVRO and Pico Veleta. The data products of the first 2.5 years can be publicly accessed at {\tt www.mpifr.de/div/vlbi/fgamma}. The cross-correlation of the Effelsberg/Pico sample with the LBAS list \citep{abdo2009} showed that 47\% of our sources are detected by {\sl Fermi}-GST. A $\chi^2$ test shows that practically all our sources are variable at all frequencies at a confidence level of 99.9\,\%. The variability amplitude (as parameterized by the modulation index $m=rms/<S>$ ) increases with frequency as expected from frequency flare evolution arguments. For Effelsberg, the calibration uncertainties are kept to the level of $3-5\,\%$ at high frequencies ($\nu\ge 23.05$\, GHz) and of $0.5 - 1\,\%$ at the low frequencies ($\nu<23.05$\, GHz). The characteristic time scales present in the light curves vary from weeks to years while there is a wealth of variability patterns both in the lightcurves and in the spectra. The findings are summarized in Fuhrmann et al. (in prep.), Richards et al. (in prep.) and Angelakis et al. (in prep.). Given the limited fraction of our sources in the LBAS list we are currently compiling an updated sample. 

%\subsection{}   %%% Second level section head (remove "%" symbol)
%\subsubsection{}   %%% Lowest level section head (remove "%" symbol)
%\section*{}    %%% Unnumbered top level section head (remove "%" symbol)
%\subsection*{}   %%% Unnumbered second level section head (remove "%" symbol)

\acknowledgements %%% Text of acknowledgements runs on after this command.
Based on observations with the 100-m telescope of the MPIfR (Max-Planck-Institut f\"ur Radioastronomie) at Effelsberg.

%%% THE BIBLIOGRAPHY
%%%
%%% CONSULT SECTION 3 OF "INSTRUCTIONS FOR AUTHORS" FOR HOW TO USE NATBIB.
%%% AUTHORS ARE ENCOURAGED TO USE EITHER THE "THEBIBLIOGRAPY" ENVIRONMENT
%%% BY UNCOMMENTING (DELETING THE "%" SYMBOL) THE COMMANDS BELOW, OR BY
%%% USING THE BIBTEX ENVIRONMENT. TO FIND OUT WHICH IS APPLICABLE TO YOUR
%%% CONTRIBUTION, CONSULT THE VOLUME EDITORS FOR YOUR PROCEEDINGS.
%%%

\end{document}